\documentclass[aip,floats,twocolumn,noshowpacs,superscriptaddress]{revtex4}
\usepackage{graphicx,epsfig}
\usepackage{times}
\usepackage{graphics,dcolumn,bm,fleqn,epic,eepic,float}
\usepackage{amssymb,amsmath,multirow,rotate,color}

\begin{document}
\title{Evolutionary Games defined at the Network Mesoscale: The Public
  Goods game}

\author{Jes\'us G{\'o}mez-Garde\~{n}es}

\affiliation{Departamento de Matem\'atica Aplicada, Universidad Rey
  Juan Carlos, 28933 M\'ostoles, Madrid, Spain}

\affiliation{Institute for Biocomputation and Physics of Complex
  Systems (BIFI), Universidad de Zaragoza, 50018 Zaragoza, Spain}

\author{Miguel Romance}

\affiliation{Departamento de Matem\'atica Aplicada, Universidad Rey
  Juan Carlos, 28933 M\'ostoles, Madrid, Spain}

\author{Regino Criado}  

\affiliation{Departamento de Matem\'atica Aplicada, Universidad Rey
  Juan Carlos, 28933 M\'ostoles, Madrid, Spain}

\author{Daniele Vilone}

\affiliation{Grupo Interdisciplinar de Sistemas Complejos (GISC),
  Departamento de Matem\'aticas, Universidad Carlos III, 28911 Legan\'es,
  Madrid, Spain}

\author{Angel S\'anchez}

\affiliation{Institute for Biocomputation and Physics of Complex
Systems (BIFI), Universidad de Zaragoza, 50018 Zaragoza, Spain}

\affiliation{Grupo Interdisciplinar de Sistemas Complejos (GISC),
  Departamento de Matem\'aticas, Universidad Carlos III, 28911 Legan\'es,
  Madrid, Spain}

\affiliation{Instituto de Ciencias Matem\'aticas CSIC-UAM-UC3M-UCM, 28049 
Cantoblanco, Madrid, Spain}

\date{\today}

\begin{abstract}
The evolutionary dynamics of the Public Goods game addresses the
emergence of cooperation within groups of individuals. However, the
Public Goods game on large populations of interconnected individuals
has been usually modeled without any knowledge about their group
structure. In this paper, by focusing on collaboration networks, we
show that it is possible to include the mesoscopic information about
the structure of the real groups by means of a bipartite graph. We
compare the results with the projected (coauthor) and the original
bipartite graphs and show that cooperation is enhanced by the
mesoscopic structure contained. We conclude by analyzing the influence
of the size of the groups in the evolutionary success of cooperation.
\end{abstract}

\maketitle

{\bf Evolutionary game dynamics on graphs has become a hot topic of
  research during the last years. The attention has been mainly
  focused on $2$-players games, such as the Prisoner's Dilemma game,
  since the pairwise interactions can be easily implemented on top of
  networked substrates. However, for $m$-players game, such as the
  Public Goods game, the microscopic description about the pairwise
  interactions contained in the network is not enough, since
  $m$-players game are intrinsically defined at the mesoscopic network
  level. This mesoscopic level describes how individuals engage into
  groups where the Public Goods games are played. However, the actual
  group structure of networks has not been considered in the
  literature, being automatically substituted by a fictitious one. In
  this work, we study the emergence of cooperation in collaboration
  networks, by incorporating the real group structure to the
  evolutionary dynamics of the Public Goods game. Our results are
  compared with those obtained when the mesoscopic structure is
  ignored. We show that cooperation is actually enhanced when the
  group structure is taken into account, thus providing a novel
  structural mechanism, relying on the mesoscale level of large social
  systems, that promotes cooperation. Moreover, we further show that
  the particular characteristics of the group structure strongly
  influence the survival of cooperation.}

\section{Introduction}

Evolutionary game theory on graphs is attracting lately a lot of
interest among the community of physicists working on complex systems
\cite{szabo:2007,roca:2009a}. This is a very appealing research topic because it
combines two important ideas. First, interactions take place on a
(possibly complex) network \cite{newman:2003,boccaletti:2006},
generalizing the lattice perspective; and, second, that the dynamics
taking place on that substrate needs not be the traditional one, but
rather it can arise from an evolutionary approach
\cite{nowak:2006a}. On the other hand, from the applications
viewpoint, studying evolutionary games on graphs is one of several
avenues proposed to understand the emergence of cooperation in
different contexts \cite{nowak:2006b}. This is a most relevant issue
that arises, for instance, in understanding the origin of
multicellular organisms \cite{maynard-smith:1995}, of altruistic
behavior in humans and primates \cite{kappeler:2006}, or the way
advanced animal societies work \cite{wilson:2000,henrich:2004}, to
name a few.

Research on evolutionary game theory on graphs focused on the problem
of the emergence of cooperation has considered mainly the Prisoner's
Dilemma game \cite{rapoport:1966,axelrod:1984}. The Prisoner's Dilemma
game (PDG) describes a situation in which cooperation is hampered by
the players' temptation to defect (defecting yields more payoff than
cooperating when facing a cooperator) and by the risk arising from
cooperation (cooperating with a defector yields the lowest payoff)
\cite{macy:2002}. This leads to a social dilemma in so far as when
players cooperate both the total benefit and the individual benefit are higher than 
when mutual defection occurs. While
evolutionary dynamics leads all the individuals to defection when
interactions take place in a well-mixed population (every player
interacts with every other one), the existence of a network
structuring the population can sometimes promote the emergence of
cooperation \cite{nowak:1992}, but this depends strongly on the
details of the network and the dynamics \cite{roca:2009a,roca:2008}.

Much less attention has been paid to the $m$-player generalization of
the PDG, also called Public Goods Game (PGG) \cite{kagel:1995}:
Cooperators contribute an amount $c$ (``cost'') to the public good;
defectors do not contribute. The total contribution is multiplied by
an enhancement factor $r<m$ and the result is equally distributed
between all $m$ members of the group. Hence, defectors get the same
benefit of cooperators at no cost, i.e., they free-ride on the
cooperators' effort. This is an alternative view of the social dilemma
posed by the so-called tragedy of the commons \cite{hardin:1968}. As
with the PDG, the evolutionary outcome of the PGG differs if played on
a well-mixed population (where once again defection is selected) or on
a network structure. Thus, Brandt {\em et al.}\ \cite{brandt:2003}
showed that local interactions can promote cooperation in the sense
that full cooperation is obtained for values of $r$ well below the
critical value $r=m$. This result, arising from simulation in an
hexagonal lattice, was later generalized to other lattices in
\cite{hauert:2003} and to scale-free \cite{barabasi:1999} graphs in
\cite{santos:2008}.

In this work, we go beyond those first results and focus on the
mesoscopic structure of the networks and relate it to the situation
represented by a PGG. Applications of this game arise naturally when a
number of people have to work together towards a common goal, either
to obtain some benefit or to avoid some negative effects. While trying
to stabilize the Earth's climate is a dramatic example of the latter
\cite{milinski:2006}, co-authoring scientific papers is a direct
application of the PGG in the positive sense. This provides a specific
setting in which we can test the ideas about the emergence of
cooperation in PGG on real social networks, as several collaboration
networks have been mapped and are publicly available
\cite{newman:2001a,newman:2001,newman:2004}. In this respect, it is
worth noticing that the asymptotic behavior of evolutionary games in
real social networks can be very different from that observed in model
networks, and, in fact, mesoscopic scales, clustering and motifs have
been shown to play a key role in governing the game dynamics
\cite{lozano:2008,lozano:2008a,assenza:2009}. Therefore, it is
important to assess to which degree, if at all, is cooperation
promoted in PGG on collaboration networks. On the other hand, while
collaboration networks are in fact bipartite, as co-authors are
connected to papers, they are very often used in a projected mode, by
connecting directly co-authors among themselves. Thus, the question
arises as to the relevance of the mesoscopic structure (as defined by
the papers) and the possible differences it may give rise to when the
original bipartite or the projected network are considered.

To study these issues in depth, this paper is structured as follows:
In section \ref{sec:model} we introduce the usual formulation of the
PGG on complex netwoks and present the new formulation based on
bipartite graphs incorporating the interaction groups, {\em i.e.} the
mesoscale structure of the population. Besides, in this section we
briefly introduce two different versions of the PGG and the different
evolutionary rules we will use. In section \ref{sec:res1} we focus on
scientific collaboration netwoks to show the different evolutionary
outcomes of the projected and the bipartite representations. Namely,
we show that the mesoscopic structure composed of the interaction
groups plays a relevant role in the promotion of cooperation. In
section \ref{sec:res2} we focus on the structural characteristics of
the mesoscale. In particular, we analyze the role of the size of the
interaction groups. The general conclusion of our analysis is that the
larger the interaction groups are, the more difficult cooperation is
promoted. Finally, in section \ref{sec:conc} we summarize the main
results and pose some relevant questions that arise from them.

\section{Modelling Evolutionary Dynamics of Public Goods Game}
\label{sec:model}

\subsection{The Evolutionary Public Goods Game}

The classical setting of a PGG models an economic or social group of
$m$ agents whose strategies can be cooperation (C) or defection
(D). As explained above, if an agent cooperates she invests a quantity $c$ into the public
pot whereas defectors do not contribute. Therefore, in a group with
$x$ cooperators (and $m-x$ defectors) the total amount of investments
is $xc$. This amount is then multiplied by an enhacement factor $r>1$
so that the total investment increases to $rxc$. This amount is then
distributed among all the participants of the PGG regardless of their
contributions. Therefore, the benefit of each defector will be
\begin{equation}
f^{D}=\frac{rxc}{m}\;,
\end{equation}
while for a cooperator the benefit decreases to $f^C=f^D-c$. From
these benefits it is obvious that defectors will earn more than
cooperators, $f^D\geq f^C$. Morevover, while $f^{D}\geq 0$,
cooperators only have possitive benefits when $rx>m$. This means that
a lonely cooperator playing with a group of defectors will always lose
($f^{C}<0$) whenever $r<m$. Therefore, the Nash equilibrium of a PGG
with $r<m$ is a full defection situation ({i.\ e.}, a group in which
all players defect). However, this equilibrium is not Pareto optimal
since full defection yields zero total reward whereas if everyone
contributes to the PGG (full cooperation) the group will obtain the
maximum total reward. Thus, here is where the social dilemma lies.

In an evolutionary context individuals are not considered fully
rational, so that they do not play a Nash equilibrium found from a
rational analysis of the PGG. Besides, agents are not organized into a
single group but in general a large population of $N\gg m$ agents is
allowed to organize into a large number of groups with $m$ agents. The
relevant difference with the classical setting is the introduction of
a dynamical evolution: Agents play the game several times and they are
allowed to change their strategy after each round of the game. These
strategy changes obey certain evolutionary rules by which agents
evaluate their performance comparing their fitness with those of the
rest of the population (see section \ref{sec:update}).

In a well mixed population the agents play within several groups
during the different rounds of the game. In particular, before each
round of the PGG the groups are formed randomly. Under this
well-mixing assumption it can be shown that the evolutionary dynamics ends
up in full defection whenever $r<m$. Therefore, defection again
dominates over cooperation as in the (static) classical
setting. Driven by the abundance of examples in which cooperation is
observed in social, economic and biological situations similar to that
defined by the PGG, it is clear that some mechanisms beyond
irrationality and evolution are at the core of the survival of
cooperation. In this line several mechanisms have been proposed such
as costly punishment \cite{brandt:2003,helbing:2010njp}, meaning the
possibility of punishing defectors after a round of the PGG, or the
addition of reputation \cite{brandt:2003} to agents, which signals the
behavior of these players in past rounds of the game. These
social-based mechanisms allow to enhance the contributions to the PGG,
thus favouring the survival of cooperative behaviors.

\subsection{The Public Goods Game on Complex Networks}
\label{pggnet}

The aforementioned mecanisms (punishment and reputation) are clearly
based in human behaviors that are plausible to appear in social
systems. However, cooperation in the PGG can also be promoted by
taking into consideration the structure of interaction between
players. To this aim, one leaves out the well-mixing assumption and
works with a static substrate of interactions. As introduced above, in
\cite{brandt:2003} it was shown that in the case of the PGG,
cooperation was significantly promoted when considering Euclidean
lattices. The importance of the structure of interactions in the
success of cooperation in the general context of evolutionary game
theory is underlined by the term {\em network reciprocity} \cite{nowak:2006b}.

\begin{figure}[!t]
\epsfig{file=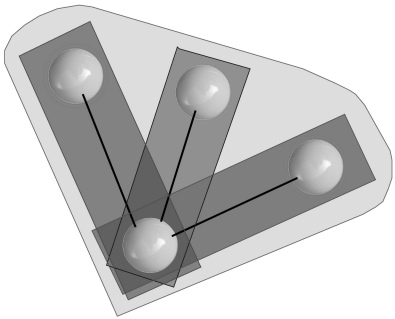,width=0.60\columnwidth,angle=-180,clip=1}
\caption{Schematic representation of the usual way for defining the
  interaction groups of the PGG on complex networks. Each of the $4$
  agents defines a group composed of herself and its neighbors. As a
  result, the above graph contains $3$ groups composed of $2$ nodes
  and a big one containg the $4$ nodes of the graph.}
\label{fig0}
\end{figure}

The interaction backbone of real social systems is however far from
Euclidean structures. In particular, many studies in the last decade
have addressed the characterization of such social systems as complex
networks \cite{newman:2003,boccaletti:2006}. These networks are a
collection of $N$ nodes (accounting for each agent of the system) and
$L$ links (describing the interaction between pairs of
agents). Complex networks typically display structural patterns that
are absent in regular geometries, such as the small-world property
\cite{watts:1998} or scale-free patterns for the number of connections
of the agents \cite{barabasi:1999}. On the other hand, real complex
networks are sparse ($L\sim N$) meaning that the well mixed assumption
(which would imply that $L\sim N^2$) does not hold. Thus, it is
necessary to study how the structure of these networks affects the
evolution of cooperation. As in the case of regular lattices, the
first evolutionary social dilemma to be studied on top of networks was
the PDG. The main result of these studies is that, under certain
conditions \cite{roca:2008,roca:2009a}, cooperation is further
enhanced whith respect to the case of regular lattices. Moreover, it
was observed that the degree-heterogeneity of scale-free networks
increases significantly the survival of cooperation with respect to
random complex networks \cite{santos:2005,gomezgardenes:2007},
as was subsequently shown \cite{santos:2008} for PGG, thus
reinforcing the message that scale-free structures are natural
promoters of cooperation.

The implementation of the PGG on top of complex neworks is however,
not as straighforward as in the case of the PDG. The reason is clear:
While the PDG is defined for pairwise interactions and thus the
possible games are dictated by the collection of links of the network,
for $m$ player groups ($m>2$) we do not have the information about how
to engage players in groups. Therefore, some {\em a priori}
assumptions about the inner group structure of complex networks have
to be made. Most of the works in the literature about the PGG on
networks assume that a complex network automatically defines $N$
different groups of players. Namely, each of these groups is defined
by considering one agent $i$ and her $k_i$ neighbors as dictated by
the network topology (see Figure \ref{fig0}). Obviously, the size of
these groups is not regular since the number of neighbors each agent
has can fluctuate around the average connectivity of the complex
network. In scale-free networks these fluctuations diverge since the
probability of finding an individual with $k$ neighbors follows a
power-law, $P(k)\sim k^{-\gamma}$ with $2\leq \gamma\leq 3$. Thus, one
finds a large number of small size groups (those centered around
agents with small connectivity) and a few of them composed of many
agents (corresponding to groups formed around the hubs of the
system). On the other hand, since each individual $i$ participates in
$k_{i}+1$ groups, the hubs participate in a large number of groups.

Given the above definition for the group structure the implementation
of the evolutionary PGG is as follows. At each time step of the
evolutionary dynamics, each player $i$ plays the PGG within the
$k_{i}+1$ groups she belongs to (using the same strategy in each
PGG). Once all the games are played, each agent $i$ collects the total
benefit, $f_i$, obtained. If the agent plays as cooperator she pays a
cost, $c_{i}$, for participating in each of the $k_{i}+1$ groups. Here
we will consider two situations for assigning the value of the
investment made in each of the PGG she participates (as introduced in
\cite{santos:2008}). First, we consider that a cooperator agent pays a
fixed cost $c_{i}=z$ per game (FCG) played; thus, her total investment
raises to $(k_{i}+1)z$. The second option is to assume a fixed cost
$z$ per individual (FCI) playing as cooperator. Therefore, in this
latter scenario, the quantity $z$ is equally distributed by
contributing a quantity $c_{i}=z/(k_{i}+1)$ to each group she
participates.

Having in mind the above two settings for deciding the contributions
of cooperator players, we can write the benefits of each agent given
her strategy and those of her first and second neighbors. If we denote
by $x_{i}^t$ the strategy of agent $i$ during round $t$ of the
PGG, so that $x_{i}^t=1$ when playing as cooperator and $x_{i}^t=0$ if
defecting, the benefit $f_{i}(t)$ obtained after the round reads
\begin{eqnarray}
f_{i}(t)&=&\sum_{j=1}^{N}A_{ij}\frac{r(\sum^{N}_{l=1}A_{jl}x_{l}^tc_l +x_j^tc_j)}{k_{j}+1}
-k_{i}x_{i}^tc_{i}
\nonumber
\\
&+&\frac{r(\sum_{j=1}^N A_{ij}x_{j}^tc_j +x_{i}^tc_i)}{k_{i}+1}-x_{i}^tc_{i}\;.
\label{netpayoff}
\end{eqnarray}
In the above equation we have made use of the adjacency matrix of
network whose entries are $A_{ij}=A_{ji}=1$ when $i$ and $j$ are
connected and $A_{ij}=0$ otherwise, with $A_{ii}=0$ (no self-links). Note that the first two terms of
Eq. (\ref{netpayoff}) correspond to the PGG played within the groups
formed around the neighbors of $i$ while the last two terms account
for the game played by $i$ and her neighbors.


\begin{figure*}[!t]
\epsfig{file=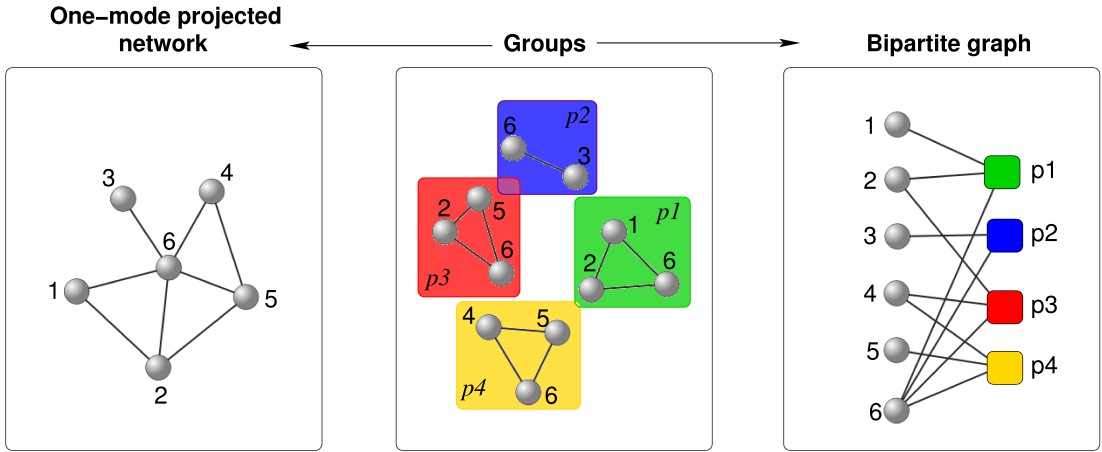,width=1.8\columnwidth,angle=0,clip=1}
\caption{Schematic representation of the two different forms of
  encoding collaboration data. In the central plot several
  collaboration groups represent the original data. The interactions
  among agents can be translated into a projected complex network
  (left). However, if one aims at preserving all the information about
  the group structure, a representation as a bipartite graph (right)
  is more appropriate.}
\label{fig1}
\end{figure*}

\subsection{The Public Goods Game on Bipartite Graphs}
\label{pggbip}

The definition of the groups where the PGG takes place as the sets
formed by each agent and her network neighbors arises from using the
network of contacts as the map of agent interactions. However, most
social networks are constructed from real data containg information
about the groups formed by individuals. Well-known examples are
collaboration networks in which agents can be scientists collaborating
to perform research \cite{newman:2001,newman:2001a,newman:2004}. In
Figure \ref{fig1} we plot how collaboration data are usually collected
to form a projected (or one-mode) complex network of the interactions
among agents (the co-author network). The central plot corresponds to
the original data containing several collaboration groups among six
agents. These groups are then translated into a complex network by
projecting the original data (left plot). The collection of groups
then transforms into a star-like graph in which there is a central hub
(node $6$) with five neighbors, some of them connected and thus
forming triangles with the central hub. One easily realizes that
groups defined on the projected network itself (following the
definition given in section \ref{pggnet}) are rather different from
the original ones.

On the other hand, one can take advantage of the information available
in collaboration data by constructing a bipartite graph
\cite{ramasco:2004,guimera:2005,peltomaki:2006}. The structure of this
bipartite graph is represented in the right plot of Figure
\ref{fig1}. As observed, the bipartite representation contains two
types of nodes denoting agents (left column of round nodes) and
collaborations (right column of squared nodes) respectively. It is
clear that connections are restricted to link nodes of different types
({i.\ e.} belonging to different columns). Thus, such a bipartite
representation preserves the information about the group structure of
the original data and constitutes a well-suited framework for studying
dynamical processes intrinsically defined at a system mesoscale
\cite{GomezGardenes:2008} (in our case defined by the collaboration
groups) as is the case of the PGG.

Let us now formalize the bipartite graph in which the evolutionary
dynamics of the PGG takes place. The graph will be composed of $N$
agents playing the PGG within $P$ groups. The particular way agents
engage into groups will be encoded by a $P\times N$ matrix $B_{ij}$
usually called biadjacency matrix. Given the bipartite structure of
the graph, the $i$-th row accounts for the individuals participating
in group $i$, so that agent $j$ is engaged in group $i$ whenever
$B_{ij}=1$ while $B_{ij}=0$ when she is absent (note that now $B_{ii}$
needs not be zero as rows and columns represent different entities). Alternatively, the
$i$-th column contains the information about the groups containing
agent $i$:  $B_{ji}=1$ when agent $i$ participates in group $j$ and
$B_{ji}=0$ otherwise. Given the biadjacency matrix we can calculate
the number of groups agent $i$ takes part, $q_{i}$, as
\begin{equation}
q_{i}=\sum_{j=1}^{P}B_{ji},\;\;\; (i=1,...,N)\;. 
\end{equation}
Alternatively, the number of participants contained in group $i$,
$m_{i}$, reads 
\begin{equation}
m_{i}=\sum_{j=1}^{N}B_{ij},\;\;\; (i=1,...,P)\;. 
\end{equation}

Having introduced the structure of the bipartite graph describing the
relations between agents and groups, we model the PGG. At each time
step, each player $i$ ($i=1,...,N$) plays a round of the PGG each at
every group she participates in as defined by the biadjacency matrix
of the bipartite graph, $B_{ji}=1$ ($j=1,...,P$). Obviously, the
benefit obtained by the agent depends on both her strategy and those
of the agents participating in the same groups. The net benefit after
playing round $t$ of the PGG now reads
\begin{equation}
f_{i}(t)=\sum_{j=1}^{P}\frac{rB_{ji}}{m_{j}}\left[\sum^{N}_{l=1}B_{jl}x_{l}^tc_l \right]-x_{i}^tc_{i}q_{i}\;.
\label{bippayoff}
\end{equation}
Note that the sum in the above expression accounts for $q_{i}$ PGGs
played by $i$ while the last term is for the cost associated to
partipating as cooperator.

\subsection{Strategy update: Evolutionary Dynamics}
\label{sec:update}

After a round of the PGG is played, agents update their
strategies. This update is driven by the benefits obtained by the
agent and her neighbors in the last round of the game. Thus, the
update stage keeps the local character, by restricting the information
available to agents about the benefits of other players to their local
(one-mode) network neighborhoods. Note that the group structure
described in the bipartite representation plays no role in this stage,
as update rules make use of the network of contacts. Thus, the update
process takes place in the same way regardless of the representation
(one-mode network or bipartite graph) of the PGG we are using.

In this work we will use three different update rules in order to test
the robustness of the results obtained. In all the update rules each
agent decides to use the strategy of a given neighbor $j$ in the next
round of the game ($x_{i}^{t+1}=x_{j}^t$) or to stay the same
($x_{i}^{t+1}=x_{i}^t$). The three update rules work as follows:
\begin{itemize}
\item Unconditional Imitation (UI) \cite{nowak:1992}: agent $i$
  compares her payoff with her neighbor with the largest payoff, say
  agent $j$. Agent $i$ will copy the strategy of agent $j$ provided
  $f_{i}<f_{j}$. Otherwise, agent $i$ will remain unchanged. The
  probability of copying agent $j$ is given by
\begin{equation}
P_{j}=\Theta(f_{j}-f_{i})\;\;{\mbox{with}}\;\; f_{j}=\max\{f_{l}|A_{il}=1\}\;,
\end{equation} 
where $\Theta(x)$ is the Heaviside step function, $\Theta(x)=1$ when
$x>0$ and $\Theta(x)=0$ for $x\leq 0$.
\item Fermi rule \cite{szabo:1998,traulsen:2006a}: agent $i$ chooses
  one neighbor at random, say agent $j$, and compares their respective
  benefits. The probability that $i$ copies the strategy
  of the chosen neighbor obeys a saturated Fermi function of the
  benefit difference $f_{i}-f_{j}$. Thus, the probability that $i$
  decides to take the strategy of an agent $j$ reads
\begin{equation}
P_{j}=\frac{A_{ij}}{k_{i}}\cdot\frac{1}{1+{\mbox e}^{\beta(f_{i}-f_{j})}}\;.
\end{equation}
\item Moran rule (MOR) \cite{kirchkamp:1999,roca:2008}: agent $i$
  chooses one of her neighbors proportionally to her
  payoff. Subsequently, agent $i$ adopts automatically the state of
  the chosen neighbor. Therefore, the probability of choosing agent
  $j$ is given by
\begin{equation}
P_{j}=\frac{A_{ij}f_{j}}{\sum_{l=1}^{N}A_{il}f_{l}}\;.
\end{equation}

\end{itemize}
These three update rules contain different evolutionary
ingredients. In particular, UI and MOR use global knowledge about the
benefits of the neighbors since they evaluate all of them. On the
contrary, the Fermi update chooses one neighbor randomly. Concerning
the stochastic character of the agent's decisions, we note that both
Fermi (for small and moderate values of $\beta$) and MOR updates are
purely stochastic and they even allow mistakes, {\em i.e.} it is
possible to copy the strategy of a neighbor with smaller benefit. In
contrast, UI is purely determistic and errors are not admitted. Note
that when $\beta\gg 1$ (strong selection limit) the saturated Fermi
function turns into a Heaviside step function thus mimicking the
behavior of UI. However, the differences in the degree of knowledge
about neighbors of both setting persist. In the following, we will use
$\beta=1$ for the Fermi update since the results are quite robust
around this value.


\section{Cooperation in Scientific Collaborations: Projected versus Bipartite networks}
\label{sec:res1}

\begin{figure}[t!]
\epsfig{file=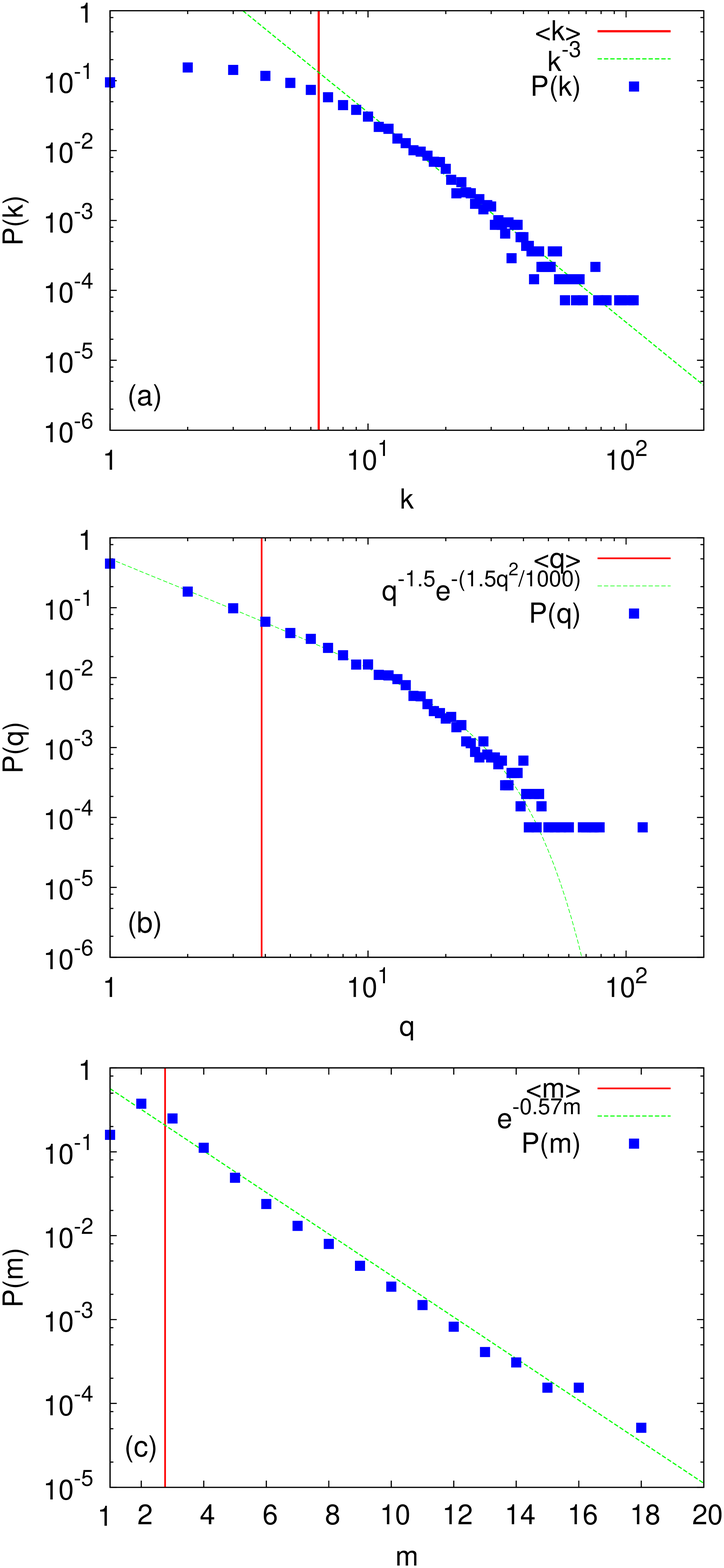,width=0.80\columnwidth,angle=0,clip=1}
\caption{Structural analysis of the {\em cond-mat} scientific
  collaboration network. In {\bf (a)} we plot the degree distribution,
  $P(k)$, of the projected (one-mode) network (coauthor network). This
  distribution display a long tail decaying as $P(k)\sim k^{-2}$. The
  average connectivity is $\langle k\rangle=6.44$ as indicated (in this and 
  subsequent plots) by a red vertical line. Plots {\bf (b)} and
  {\bf (c)} show the characteriztic of the associated bipartite
  graph. In {\bf (b)} we show the degree distribution of authors,
  $P(q)$, {\em i.e.} the probability of finding and author
  contributing to $q$ papers. The behavior of this distribution
  denotes a sharp decay thus indicating that the initial power law
  behavior truncates for large $q$. The average number of papers per
  author is $\langle q\rangle=3.87$. On the contrary, in {\bf (c)} we
  plot the probability that a paper is coauthored by $m$ authors,
  $P(m)$. In this case $P(m)$ decays exponentially (note normal scale
  on $x$ axis) and, on average,
  papers are coauthored by $\langle m\rangle=2.76$ researchers.}
\label{scn}
\end{figure}

In this section we implement the PGG on top of a real collaboration
network. The network is composed by $N=13861$ scientist and the
collaboration data is obtained from $P=19465$ papers appeared in the
{\em cond-mat} section of the arXiv preprint server \cite{newman:2001a}. This collection
of papers is obtained after computing the giant connected component of
the (projected) coauthor network of the original data set
 which has $16726$ authors and $22015$ papers. In
Figure \ref{scn} we plot the degree distribution of the coauthor
network and those of the bipartite (authors-papers) graph. Both the
probability of finding one author with $k$ coauthors, $P(k)$, and that
of having an author collaborating in $q$ papers, $P(q)$, have broad
profiles. On the other hand, the probability that a paper is
coauthored by $m$ researchers, $P(m)$, shows an exponential
decay. This homogeneous distribution for the number of authors
coauthoring one paper is a very important difference arising when
comparing the (one-mode) coauthor network with the bipartite
representation of the collaboration data.

The structural differences between the coauthor network and the
bipartite graph imply that the dynamical processes implemented on top
of them can yield different results. In particular, modelling the PGG
without any knowledge of the real group structure will give as a
result the definition of large groups centered around hubs of the
coauthor network (see Figure \ref{scn}.a). However, this definition
strongly contrasts with the homogenenous distribution $P(m)$ for the
number of authors collaborating in one paper. Thus, we will compare
the outcome of the PGG evolutionary dynamics using the one-mode
coauthor network as in \cite{santos:2008} with the results obtained by
working with the real collaboration data, {\em i.e.} with the
bipartite graph, in which the group structure arises in a natural manner
as defined by the set of papers.

We will focus on the evolution of the asymptotic value of the
cooperation level, $\langle c\rangle$, as a function of the enhacement
factor $r$. The cooperation level usually represents the fraction of
the $N$ individuals that cooperate in the stationary regime. Thus, in
our simulations we start by assigning randomly the initial strategies
of the players, $\{x_{i}^{0}\}$, so that half of the populations plays
initially as cooperators and the other one as does as defectors.
Then, we let the evolutionary dynamics evolve for $\tau=10^5$ rounds
of the PGG and measure the stationary value of the cooperation level
during $T=10^4$ additional rounds. Thus, the final value of $\langle
c\rangle$ is computed as
\begin{equation}
\langle c\rangle =\frac{1}{T\cdot
  N}\left(\sum_{t=\tau+1}^{\tau+T}\sum_{i=1}^Nx^{t}_{i}\right)\;.
\label{cmed}
\end{equation}
The above definition of $\langle c\rangle$ assummes that the
evolutionary dynamics ends up in a dynamical equilibrium in which
cooperators and defectors coexist. However, for the Fermi and MOR
updates, depending on the precise values of $r$, this is not the
case. Quite on the contrary, each run of the evolutionary dynamics for
the same value of $r$ (corrresponding to a different set of initial
conditions) ends up into either full defection or full
cooperation. The strong ergodicity of the dynamics, due to the
stochastic character of these updates rules, drives the system
evolution into one of those two absorbing states. Therefore, it is
mandatory to perform a large number (at least $10^3$ in our case) of
different realizations (corresponding to different initial conditions)
of the evolutionary dynamics. Obviosuly, in those cases where the
dynamical evolution always finishes in one of the two absorving
states, the reported value of $\langle c\rangle$ is defined as the
fraction of realizations in which the dynamics ends up in full
cooperation.

\begin{figure*}[!t]
\epsfig{file=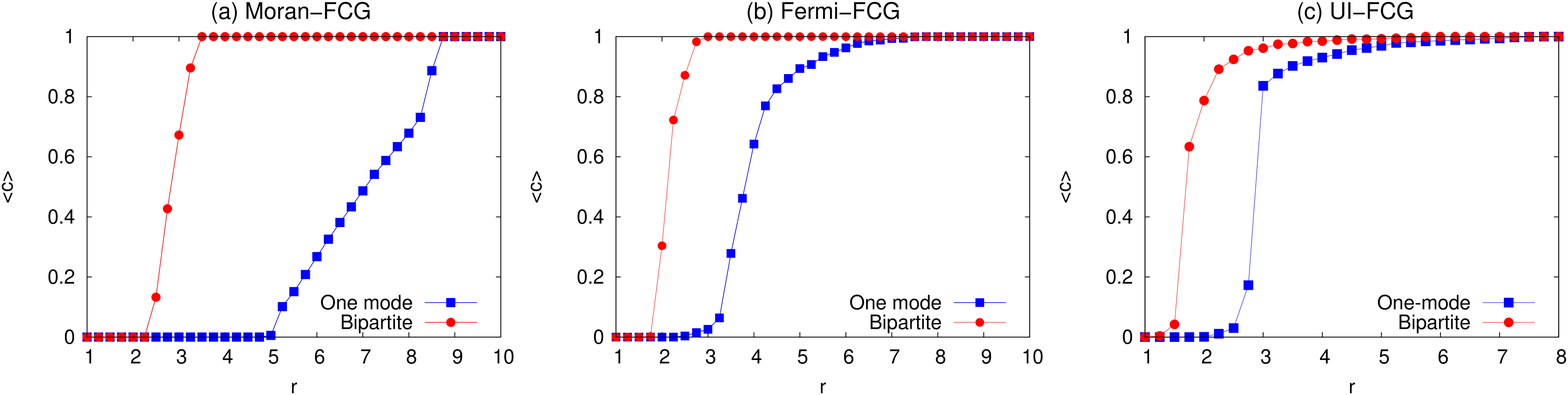,width=1.8\columnwidth,angle=0,clip=1}
\\
\epsfig{file=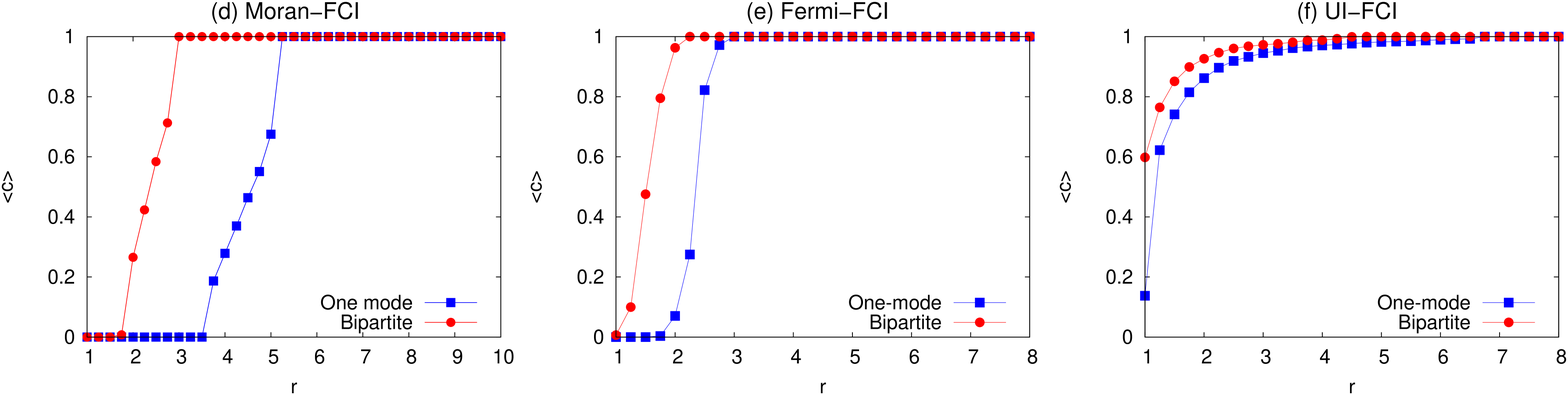,width=1.8\columnwidth,angle=0,clip=1}
\caption{Cooperation level $\langle c\rangle$ as a function of the
  enhancement factor $r$ for the PGG played on top of the one-mode
  (projected) coauthor network and the bipartite graph preserving the
  original group structure. The first three plots {\bf (a)}, {\bf (b)}
  and {\bf (c)} correspond to the PGG with fixed cost per game (FCG) while the plots {\bf
    (d)}, {\bf (e)} and {\bf (f)} account for the PGG with fixed cost per
    individual (FCI). For
  each of the two versions of the PGG we show the evolution of the
  curves $\langle c\rangle (r)$ for three different update rules: {\bf
    (a)} and {\bf (d)} MOR update, {\bf (b)} and {\bf (e)} Fermi rule,
  and {\bf (c)} and {\bf (f)} UI.}
\label{sci}
\end{figure*}

Figure \ref{sci} shows the function $\langle c\rangle (r)$ for both
the (one-mode) coauthor network and the bipartite graph in six
different scenarios. Namely, plots \ref{sci}.a, \ref{sci}.b and
\ref{sci}.c show the results for the PGG payed with fixed cost per
game (FCG) while in plots \ref{sci}.d, \ref{sci}.e and \ref{sci}.f we
show the case of the PGG played with fixed cost per individual (FCI).
As introduced in section \ref{sec:update}, for both the FCG and FCI
versions of the PGG we show the outcome of the evolutionary dynamics
when three update dynamics are at work. Namely, in plots \ref{sci}.a
and \ref{sci}.d we use the MOR (strongly stochastic and using
global knowledge) scheme, in plots \ref{sci}.b and \ref{sci}.e the
Fermi rule (slightly stochastic and with limited knowledge),
 and finally, plots \ref{sci}.c and \ref{sci}.f correspond to
UI update (purely deterministic and using global knowledge).

As can be seen from the plots, the average level of cooperation
$\langle c\rangle$ increases from $\langle c\rangle=0$ to $\langle
c\rangle=1$ when the value of $r$ exceeds some threshold $r_t$. The
precise value of this threshold and the velocity of this transition
depends strongly on the particular dynamical rule and the substrate of
interactions used. It is clear that our main interest here is to
confront the results of the PGG obtained using the one-mode network
and the bipartite graph. The plots corresponding to the PGG with FCG
clearly show that the cooperation level is always larger (meaning that it 
sets on for lower values of $r$ and increases faster) when the
structure of groups is that of the real collaboration data, {i.\ e.}
of the bipartite representation. It is also clear that the MOR update
rule (plot \ref{sci}.a) gives rise to larger differences between the two
substrates. Interestingly, we observe that the curve $\langle c\rangle
(r)$ corresponding to the bipartite graph is much more stable under
update rule changes than its one-mode counterpart. On the other hand,
for both the one-mode and the bipartite substrates cooperation
increases when the stochastic character of the update rule decreases,
{\em i.e.} going from MOR update to the Fermi rule and from the Fermi
rule to UI.

The FCI setting is probably the most appropriate version of the PGG to model
scientific collaborations. The reason is clear, researchers have a
limited amount of time/resources to invest in collaborating and it has
to be partitioned among all the collaborations they share. In general,
researchers participating in a large number of projects tend to
contribute less (in terms of time and lab work) to each paper in which they
appear. On the other hand, those researches involved in few
collaborations tend to assume the largest part of the work to do. In
the plots of Figure \ref{sci} corresponding to the PGG in its FCI
version we find the same result as for the PGG with FCG: the group
structure (contained in the bipartite graph) promotes
cooperation. Again, the differences between both substrates are
larger when using the MOR update while the stochasticity of the
update rules decrease the level of cooperation in both cases.

\begin{figure*}[!t]
\epsfig{file=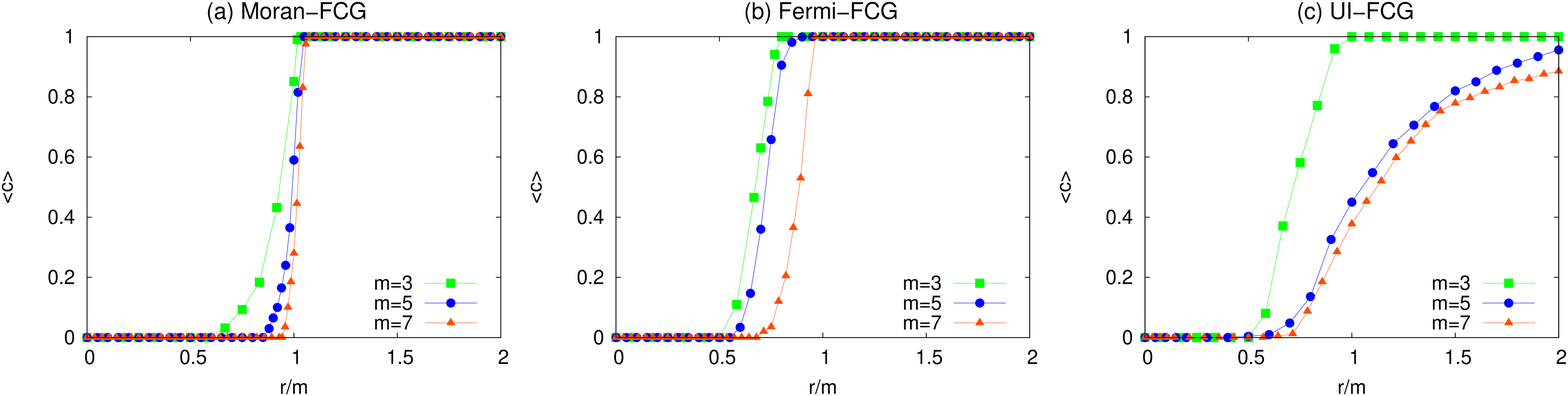,width=1.8\columnwidth,angle=0,clip=1}
\\
\epsfig{file=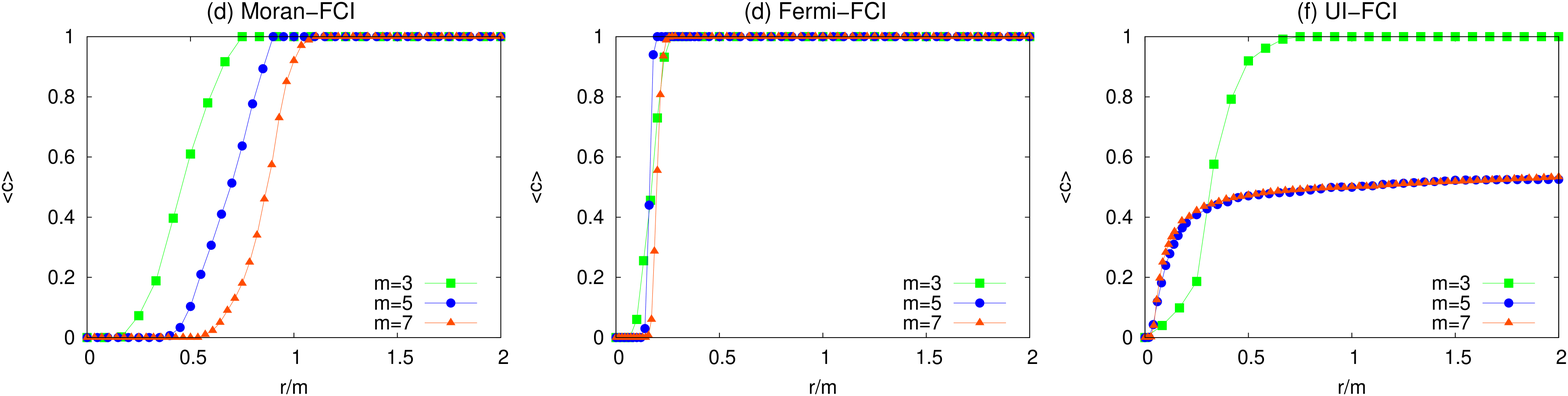,width=1.8\columnwidth,angle=0,clip=1}
\caption{Cooperation level $\langle c\rangle$ as a function of the
  rescaled enhancement factor $r/m$ for the PGG played on top of three
  synthetic collaboration networks with different group sizes $m=3$,
  $5$ and $7$. As in Figure \ref{sci}, the first three plots {\bf
    (a)}, {\bf (b)} and {\bf (c)} correspond to the PGG with FCG while
  the plots {\bf (d)}, {\bf (e)} and {\bf (f)} account for the PGG
  with FCI. For each of the two versions of the PGG we show the
  evolution of the curves $\langle c\rangle (r/m)$ for three different
  update rules: {\bf (a)} and {\bf (d)} MOR update, {\bf (b)} and {\bf
    (e)} Fermi rule, and {\bf (c)} and {\bf (f)} UI.}
\label{synt}
\end{figure*}

\section{Influence of Group Size in the Promotion of Cooperation}
\label{sec:res2}

Having shown that the mesoscopic group structure of collaboration
networks strongly affects the promotion of cooperation, we now 
consider the issue of the size of the groups in which PGG is played. 
To this end, and inspired in the model
introduced by Ramasco {\em et al.}  \cite{ramasco:2004}, we propose
the following way for constructing synthetic collaboration graphs. We
start with an initial core of $m$ nodes that defines the first group
of our bipartite graph. At each time step of the growth process we add
a new element, that will define a new group of size $m$. To do this,
the newcomer chooses one of the nodes already present in the
graph. The probability $P_{i}$ that a node $i$ receives the link from
a newcomer is proportional to the number of groups it belongs to,
$g_{i}$,
\begin{equation} 
P_{i}=\frac{g_{i}}{\sum_{j}g_{j}}\;.
\end{equation}
Once the newcomer has chosen the first node, say with node $j$, it
closes the group by choosing other $(m-2)$ nodes randomly from the
neighbors of $j$, {\em i.e.} among those nodes that participate in one
or more groups with $j$. The above process is iterated until the graph
contains $N$ nodes (and $N-m+1$ groups). The above model, being
extremely simple, allows to reproduce two main structural features
observed in collaboration networks: the scale-free distribution for
the number of contacts each individual has in the projected (coauthor)
network and the nearly constant value for the number of authors
appearing in a paper. In fact, this latter feature is used as a
tunable parameter, $m$, in our network model allowing us to explore the effect
that this size has on the evolution of cooperation. In the following
we will fix the size of the network to $N=5000$ and we will work with
$m=3$, $5$ and $7$.

Following the same strategy as in the previous section we will compare
the outcome of the evolutionary dynamics making use of three update
rules (MOR, Fermi and UI) and we will also analyze the PGG in both its
FCG and FCI versions. In Figure \ref{synt} we show the six plots
corresponding to these scenarios. The initial setup and the
numerical procedure is identical to that used in the previous
section. The only novelty is the use of the rescaled enhancement
factor, $r^{'}=r/m$, so to compare the outcome of the PGG dynamics in
different network topologies (they depend heavily on $m$)
\cite{santos:2008,szolnoki:2009a} here
labeled by the group size $m$.

Let us start by analyzing the case of the PGG played with FCG. In this
case the curves $\langle c\rangle (r/m)$ in plots \ref{synt}.a and
\ref{synt}.b, corresponding to the MOR and Fermi (stochastic) updates,
behave as expected: Cooperation dominates for $r/m>1$ ({i.\ e.},
when the enhancement factor is larger than the group size) while for
$r/m<1$ it decays fast towards full defection. The decay becomes
sharper as $m$ increases so that we conclude that small groups benefit
cooperation. The case of UI (plot \ref{synt}.c) confirms this
conclusion about the negative effects of large groups. However, in
this case the curves $\langle c\rangle (r/m)$ for $m=5$ and $7$ point
out a dramatic scenario for the survival of cooperation. While for the
rest of the curves $r/m=1$ represent the point beyond which full
cooperation dominates in those curves corresponding to UI with $m=5$
and $m=7$ the transition is very slow. Therefore, the effects of
enlarging the group size in the mesoscopic structure of collaboration
networks seem to have negative effects over cooperation, specially in
the case when UI is the update mechanism at work.
 
Now we turn our attention to the PGG played wit FCI. As before, we
first focus on the stochastic update rules (MOR and Fermi). In the
corresponding plots (\ref{synt}.d and \ref{synt}.e) we observe that,
for the same value of the group size $m$, cooperation is significantly
enhanced with respect to the case of the PGG with FCG. In the case of
the MOR update we also observe again (as in the PGG with FCG) that by
increasing the group size the cooperation level decreases. However,
for the Fermi rule this is not the case (at variance with the PGG with
FCG) and the curves $\langle c\rangle (r/m)$ collapse in the
transition region, placed around $r/m\simeq 0.2$. The case of the UI
turns to be the most intriguing as in the PGG with FCG. However, in
the case of the FCI version the effects of enlarging the size of the
groups have worse consequences as observed from the plot
\ref{synt}.f. As expected, for a group size of $m=3$ cooperation is
enhanced with respect to the FCG situation, however for $m=5$ and
$m=7$ both curves are nearly the same and the situation is completely
different to that observed for $m=3$. First, for low values of $r/m$
the cooperation levels observed for $m=5, 7$ are rather large compared
to the case $m=3$ and the other curves corresponding to different
update rules. This sudden onset of cooperation is however followed by
an extremely slow increase of the cooperation level. We have checked
the roots of this behavior by looking at the dynamical evolution of
the fraction of cooperators for several realizations of the
dynamics. The result is that, despite of the deterministic character
of UI dynamics, we observe that the dynamics behaves as in the
stochastic settings, {\em i.e.} the dynamics always ends up into full
defection or full cooperation. This convergence, at variance with the
stochastic settings, is achieved in few rounds of the PGG, thus
pointing out that the dynamical outcome is strongly dependent on the
initial conditions. For UI updates the influence of the most connected
players, here represented by those agents participating in a large
number of groups, is the key role driving the evolution of the
system. Therefore, the existence of large groups enhances both the
ubiquity of those players and their benefits. The imitation process
provides with an efficient way to spread their initial strategy and
trap the system dynamics in one of the two absorbing states.

\section{Conclusions}
\label{sec:conc}

Summarizing our main results, we have shown that it of utmost
importance to include the mesoscopic details about the real group
structure when dealing with the PGG on networks. The intrinsic group
structure (described by means of a bipartite graph) promotes
cooperation in PGGs, this being a new mechanism for this phenomenon
beyond the scale-free character \cite{santos:2008} and other features
\cite{huang:2008,rong:2009,shi:2010,rong:2010} of the one-mode
(projected) complex network.  Regarding the size of the groups in wich
the PGG takes place, we have shown that they affect the outcome of the
evolutionary dynamics in an important way: In most of the cases,
increasing the number of the participants in each of the groups leads
to a decrease of the cooperation level. However, this decrease is
influenced by the update rule used. While for MOR and Fermi updates
the influence of the size of the groups is quite soft for the case of
UI we have shown that large group sizes slow down the development of
cooperation due to the large influence of those players participating
in a large number of groups.

Our work allows us to draw important conclusions regarding the application
of these models and the corresponding research. Thus, looking again at the
difference in the behavior observed on the bipartite network and on the 
projected one, it is clear that the fact that the mean group size in both 
settings is different plays a role in the promotion of cooperation: Indeed,
as is known for PGGs, smaller group sizes require smaller values of $r$
for cooperation to become a profitable strategy. This obvious fact does not
decrease the relevance of our conclusions, because what we are showing 
is that considering a projected network leads to an overestimating of the 
amplification factor needed for cooperation, arising from the artificially 
increased group size. The results in the FCG setting demonstrate that 
amplification factors between $1$ and $2$ already lead to cooperation, 
which are reasonable values in the context we are dealing with, namely
collaboration in research and paper-writing. On the other hand, the large 
value obtained for the MOR rule indicates that this is not likely to be a 
good model of human behavior in this context, while local imitative rules
like Fermi or UI yield lower estimates for the critical $r$, probably closer
to reality. Note also that we have seen important differences between 
a setup in which the amount one can invest is unlimited (FCG) or bounded 
(FCI). This latter scenario, which is closer to reality in the sense that we all
have limited time and energy to devote to collaborative work, gives rise to
very low (or even smaller than $1$) critical values for $r$. This might seem
strange at first glance, but when considering this issue on the light of the
structure of the bipartite network, one realizes that even with the bipartite
description there are authors with a large number of collaborations, i.\ e.,
there are hubs. These hubs invest very little on every collaboration they 
are involved in, and in practice become free-riders. However, imitative 
update rules forces their neighbors to be cooperators as well, because they
observe the large payoff received by the hub (arising from his many 
collaborations), and only under MOR dynamics larger values of $r$ are
needed to support cooperation. 

On a different note,  
our research confirms the intuition that the larger teams are, 
the more difficult it becomes to foster collaborative work. This is a very 
relevant insight in so far as it can not be obtained by looking at the projected
network, where the information about group size is lost. Our simulations on 
a simple model of collaborative network lead to the prediction that, generally
speaking, group sizes around $m=3$ are best to promote cooperation. 
Note, however, that under Fermi dynamics, the group size is not that 
important, particularly in the more realistic FCI scenario, for which the 
critical value of $r$ appears to be linearly dependent on $m$, thus making
the group size lose influence. The opposite case arises when UI is used
to update strategies, showing that it might be impossible to reach full 
cooperation even for very large values of $r$.
It is then clear that accurately modeling the collaboration 
structure is a key issue when trying to understand why people work together
in small groups, with group size and the bipartite character of the network being
particularly relevant aspects. Further research is needed to ascertain the way in
which individuals update their strategies to complete this incipient modeling toolbox.

\begin{acknowledgments}
This work has been partially supported by MICINN (Spain) through
Grants FIS2008-01240 (J.G.G.), MOSAICO (D.V. and A.S.), and
MTM2009-13838 (J.G.G., M.R.\ and R.C), and by Comunidad de Madrid
(Spain) through Grant MODELICO-CM (A.S.). D.V.\ acknowledges support
from a Postdoctoral Contract from Universidad Carlos III de Madrid.
\end{acknowledgments}

\end{document}